\documentclass[conference]{IEEEtran}
\IEEEoverridecommandlockouts
\usepackage{cite}
\usepackage{amsmath,amssymb,amsfonts}
\usepackage{algorithmic}
\usepackage{graphicx}
\usepackage{textcomp}
\usepackage{xcolor}
\usepackage{verbatim}
\usepackage{enumitem}
\usepackage{enumerate}
\usepackage{multirow}
\usepackage{ragged2e}

\def\BibTeX{{\rm B\kern-.05em{\sc i\kern-.025em b}\kern-.08em
    T\kern-.1667em\lower.7ex\hbox{E}\kern-.125emX}}
\begin{document}

\title{Graph Embedding Based Hybrid Social Recommendation System \\}
\pdfoutput=1

%\begin{center}

\author{\IEEEauthorblockN{Vishwas Sathish}
\IEEEauthorblockA{\textit{Dept. of CSE } \\
\textit{PES University}\\
Bangalore, India \\
vishwassathish@gmail.com}
\and
\IEEEauthorblockN{Tanya Mehrotra}
\IEEEauthorblockA{\textit{Dept. of CSE } \\
\textit{PES University}\\
Bangalore, India \\
tanyamehrotra025@gmail.com}
\and
\IEEEauthorblockN{Simran Dhinwa}
\IEEEauthorblockA{\textit{Dept. of CSE } \\
\textit{PES University}\\
Bangalore, India \\
simrandhinwa@gmail.com}
\and
\IEEEauthorblockN{Bhaskarjyoti Das}
\IEEEauthorblockA{\textit{Dept. of CSE}\\
\textit{PES University}\\
Bangalore, India \\
Bhaskarjyoti01@gmail.com}

}

\maketitle
%\end{center}
%%%%%%%%%%%%%%%%%%%%%%%%%%%%%%%%%%%%%%%%%%%%%%%%%%%%%%%%%%%%%%%%%%%%%%%%%%%%%%%%
\begin{abstract}

Item recommendation tasks are a widely studied topic. Recent developments in deep learning and spectral methods paved a path towards efficient graph embedding techniques. But little research has been done on applying these graph embedding to social graphs for recommendation tasks. This paper focuses at performance of various embedding methods applied on social graphs for the task of item recommendation. Additionally, a hybrid model is proposed wherein chosen embedding models are combined together to give a collective output. We put forward the hypothesis that such a hybrid model would perform better than individual embedding for recommendation task. With recommendation using individual embedding as a baseline,  performance for hybrid model for the same task is evaluated and compared. Standard metrics are used for qualitative comparison. It is found that the proposed hybrid model outperforms the baseline.
\end{abstract}

\begin{IEEEkeywords}
Graph Embedding, Social networks, Recommendation System, Social graph, Spectral Clustering, hybrid models. 
\end{IEEEkeywords}

\section{\textbf{Introduction}}

Recommendation systems (RS) are an important tool in today's digital world.  With the advent of modern web technologies, big data and distributed data management techniques, recommendation systems have helped websites like Google, Amazon and Netflix to select and display items specific to a user based on their preferences and previous actions \cite{memon2010social}. Recent developments in deep learning has only reinforced the improvement in recommendation tasks \cite{zhang2019deep}.
\par
But user's social network has not be extensively used in these systems. Social influence is a very strong factor in the choices we make in our daily lives. It is more so in the movies we watch and music we listen to. Each node (also called an actor) in the social graph refers to an entity participating in the network. The associations used to describe the relationships between these entities are referred to as edges or ties \cite{wasserman1994social}.
\par
Graph embedding are primarily representations of graphs in a lower dimension on which graph algorithms can be applied effectively. These embedding have mostly become popular as a result of computational cost of performing analytics directly on large graphs we see today, with millions of nodes and billions of edges \cite{cai2018comprehensive}. The output of a graph embedding algorithm is a vector representing the graph in a lower dimension. These powerful algorithms preserve particular aspects of graph like structural equivalence or community structures. This property can be very useful when close friends (similar due to homophily) in a social graph need to be embedded near to each other in the embedding space.
\par
In this paper, we show how social influence alone can contribute towards item recommendation. YELP data set \cite{challenge2013yelp} for restaurant business is chosen for this purpose. It provides an explicit database containing users and friends from which social graph can be extracted and pre-processed. A few embedding methods among many developed during recent years are chosen for our problem and applied on the social graph. Each of the chosen methods preserve different aspects of the social graph. We compare quality of recommendations from each of these embedding methods using the standard metrics Coverage \cite{he2010social} and Mean Average Error \cite{yang2015user}. Finally, a novel deep learning based hybrid model is proposed, that takes into account the recommendations from each chosen embedding method, combines the results and recommends best items. We propose that our approach of using social networks and hybrid of graph embedding solves the cold start and sparsity problems commonly occurring in collaborative filtering based recommendation systems \cite{elahi2016survey}.

\section{\textbf{Existing Work}}

\subsection{Recommendation Systems}

Ricci et al.\cite{ricci2011introduction} defines a recommendation system (RS) or recommender engine as a sub class of information filtering system that aims to provide suggestions to a user or predict rating a user would give to an item. It explores some common challenges faced by recommendation systems like scalability of algorithm, diversity in recommendations, long and short term preferences and other privacy issues.
\par
Bobadilla et al.\cite{bobadilla2013recommender} explores the evolution of recommendation systems over time and with the information explosion we encounter in web since the 21st century. It classifies RS based on the filtering algorithm used, into 
\begin{itemize}
\item {Collaborative filtering that takes user preferences and user data before making recommendations. Once enough data is collected, it uses this to directly recommend or find users with similar data and recommend their preferences.}
\item {Content based filtering makes recommendations based on user choices made in the past. It also directly takes the items and their content that can be analyzed, to measure similarity between the items and recommend similar ones.}
\item {Demographic filtering takes into account that similar people with some similar personal attributes like age, sex or location have common preferences.}
\item {Hybrid filtering commonly uses a combination of collaborative filtering and content based filtering to exploit the advantages of each technique.}
\end{itemize}

\par
Other papers that comprehensively discuss modern recommendation systems include \cite{zhang2019deep}, \cite{nagarnaik2015prof}, \cite{almazro2010survey} and \cite{zhang2017modeling}.

\subsection{Social Recommendation}

Social recommendations mostly make use of an explicit social graph or a set of friends from the social circle of a user. This provides an alternate method for recommendation by avoiding problems faced in traditional recommendation systems such as cold start, scalability and sparsity problems seen in collaborative filtering methods. The degree of avoidance of these problems and others of course depend on the inherent algorithm used, and the extent to which use of social graph is made in the algorithm \cite{memon2010social}.
\par
Kim et al.\cite{kim2012hybrid} proposed Hybrid Recommender Systems using Social Network Analysis. They argue that the popular collaborative filtering methods have major drawbacks as mentioned above. Apart from that, another disadvantage of CF is its inability to reflect qualitative and emotional information about the user. They have come up with cluster indexing collaborative filtering, by using social network analysis to select subgroups of users from internet communities.
\par
Trust based Recommendation system by Selmi et al.\cite{selmi2016trust} is divided into five sections.
\begin{itemize}
\item {The web crawler that collects information into files.}
\item {Process raw information from the files.}
\item {Analysis component that analyses the users' static attribute and interaction between users.}
\item {Create the social networks from the information collected. }
\item {Find out the core components of the social network.}
\end{itemize}
\par
Farseev et al.\cite{farseev2017cross} introduces a new method for recommendation by utilizing both user and group knowledge while using the data from multiple social media sources. It eliminates the need of domain knowledge by creating an inter-network relationship graph based on the data and out-performs the state of the art recommendation practices.

\subsection{Graph embedding}

Hamilton et al.\cite{hamilton2017representation} talks about different views of graphs. Community structure and and Structural equivalence. Incorporating the graph structure into machine learning algorithms is the main challenge faced by programmers. Some traditional methods to do so are to tabulate graph statistics like degree and clustering coefficients. Here, a new method is suggested that is suitable for graphs having million of nodes and edges. This comprises of an encoder-decoder framework. The role of the encoder is to map the large input graph into a low dimensional feature, also known as embedding. The decoder, on the other hand, decodes structural information of the graph from the embedding representation. The embedding generated by the encoder preserved the graph information both at a local and global level. Such an embedding can help in retaining as much information of the graph as possible. 

Cai et al.\cite{cai2018comprehensive} classifies the study of graph embedding into problem setting perspective and embedding technique perspective. In the problem setting perspective, embedding applied on homogeneous graphs (user-user), heterogeneous graphs (user-item) and graphs with other information type input are studied. Embedding algorithms which output embedding for node, edge, communities and the whole graph are then investigated.
\par
Cai et al.\cite{cai2018comprehensive} and Goyal et al.\cite{goyal2018graph} classify embedding techniques based on the algorithm or technique used. These mainly include 
\begin{itemize}
\item{Matrix factorization methods including HOPE \cite{ou2016asymmetric}, spectral clustering \cite{belkin2002laplacian}, and graph factorization \cite{ahmed2013distributed}.}
\item{Deep learning and random walk methods, including DeepWalk \cite{perozzi2014deepwalk}, node2vec \cite{grover2016node2vec} and Graph convolution networks \cite{kipf2016semi}. Node2vec is particularly interesting as it derives the intuition from skip gram NLP models like word2vec \cite{mikolov2013distributed} \cite{mikolov2013efficient}.}
\end{itemize}
\par
Symeonidis et al.\cite{symeonidis2013spectral} showcases the use of spectral clustering to partition the data multi-way. It is compared to other clustering algorithms like k-means and DBSCAN, while solving the problem of link prediction for recommendation.

\par
Wen et al,\cite{wen2018network} studies one embedding method, matrix factorization and perform recommendations directly on them. The study uses Ciao and Epinions data set. Liu et al.\cite{liu2019real} performs hybrid recommendation on twitter and Last.fm data sets, applying real time dynamic embedding on heterogeneous user-item graphs using modern learning methods.

\subsection{Spectral Clustering}
Practically we see most data is non convex in nature and algorithms like k-means get trapped in local optimum. Spectral clustering \cite{ng2002spectral} does not have this problem and involves with forming clusters on an embedding that is created using the eigen structure of the affinity matrix. So it basically transforms the data in higher dimension to a lower dimension and takes the most relevant eigenvectors required just enough to form clusters. Spectral clustering aims at creating disjoint sets of vertices so that all vertices falling in the same cluster have high similarity. The basic idea aims at minimizing the cost function related to the graph structure which is the basically minimizing the sum of weights of the edges that connect various clusters together. 
\par
U. Von Luxburg \cite{von2007tutorial} starts with saying that clustering of nodes for a large graph is a very widely used technique. The basis of finding such communities reside upon an objective function. However, many problems arise from the existing ways of finding communities. They may introduce bias between clusters having equal size. To overcome this, a new objective function was proposed in this paper, which focuses on two things i.e. the sum of the weights within a cluster and the sum of the weights of all the edges that are attached to the nodes in the particular cluster. To find the optimal number of clusters ‘k’, this modularity function is computed for different ranges of value of ‘k’. The value for which the modularity function is optimized gives us the exact number of clusters into which the data should be divided. 

\section{\textbf{Data set}}
We use the famous Yelp  Data set \cite{challenge2013yelp} for restaurant business to study the recommendation task, throughout the paper. The main challenge while taking the subset of the data was to retain the explicit social network, since our recommendation system is based on social circles. Hence, the data for two cities was taken, based on the large number of active users.

To set up an environment suitable for applying embedding and recommend restaurants, we first convert a sparsely connected unweighted graph $G$ to an implicit weighted graph $G^{'}$, by assigning the weights $W_{ij}$ to randomly selected pair of active users $n_{i}$ and $n_{j}$, based on similarity between them.
\[W_{ij} = \frac{|L_{i} \cap L_{j}| + (|D_{i} \cap D_{j}|}{|L_{i} \cup L_{j} \cup D_{i} \cup D_{j}|} \]
where $L_i$, $D_i$, $L_j$, $D_j$ are the set of restaurants liked and disliked by users $i$ and $j$ respectively. This similarity score increases with increase in the number of similar liked or disliked restaurants by $i$ and $j$ and is finally normalized by total number of movies both have seen. This ensures there is no bias towards users who have rated too many restaurants.

Now we have $G'$, an implicit weighted graph built over the unweighted, explicit user graph. We can apply any state-of-the-art embedding algorithm to $G'$, to map it into a lower dimension space preserving certain properties of the graph.

\section{\textbf{Methodology}}

\subsection{Problem formulation}
Here, we formally introduce the mathematical problem being solved in this study. Given a weighted and undirected social graph $G^{'}$, with $w_{ij}$ representing the weighted edge between users $u_i$, $u_j$ and $W$ representing the adjacency matrix for $G^{'}$, and given a random user $u_x$, an actor in $G^{'}$, output a list of top $k$ items as recommendation for that user $u_x$.
The graph $G^{'}$ after pre-processing, had 14,346 nodes and 407,495 edges, with an average degree of 56.8096, which is reasonable for any online social network.

\subsection{Recommendations from individual embedding}

The idea was to map users in the graph $G^{'}$ into a $D$ dimensional space. We argue that three embedding algorithms, namely spectral embedding, HOPE and node2vec would be most effective to solve our problem. We fix $D$ to be 25 for each embedding technique as it would clearly suffice for the given graph $G^{'}$ with 14,346 users.

\begin{itemize}
\item
For Node2Vec, Grover et al.\cite{grover2016node2vec} uses sampling strategy to create embedding for a given graph. Node2vec preserves higher order proximity between the nodes since the probability of occurrence of the nodes following each other is a random walk of fixed length which is maximized. A suitable tradeoff between Breadth First Search(BFS) and Depth First Search(DFS) enables node2vec to take into consideration the structure equivalence of the nodes. It also helps maintain the structure of the community.

Here $D$ is 25 and the value of $k$, number of clusters is chosen with the help of elbow method. For Node2Vec, an optimal value for the number of clusters came out to 3.

\item{Spectral embedding}
 are directly derived by spectral clustering methods \cite{belkin2002laplacian}. Spectral embedding is used since it preserves the structural information of the graph and uses the concept of independent eigenvectors in nth dimensional space \cite{ng2002spectral} \cite{von2007tutorial}. Finally k smallest eigenvalues are chosen and k clusters can be obtained by performing k means algorithm on the eigenvectors corresponding to these eigenvalues. 

Further studies on spectral embedding and spectral graph theory is done in \cite{jiang2018spectral} and \cite{luo2003spectral}.

\item{Higher Order Proximity Embedding}
or (HOPE) \cite{ou2016asymmetric} is an embedding that is primarily used to preserve higher order proximity in a graph. Higher order proximity here represents social circles and network of friends. If $k^{th}$ neighbours of two nodes in a graph are similar (not necessarily same), they are said to have similar $k^{th}$ proximity. HOPE uses a generalized version of Singular Value Decomposition to obtain the graph embedding efficiently. 

\end{itemize}

For getting recommendations from embedding, we require two types of information - the top users list and their high rated restaurants and low rated restaurants. Top users are determined by dense connectivity of users in the graph. The high rated and low rated restaurants are found by the rating that each user has given to the restaurants he has reviewed. For each user, the main aim is to find the recommended set, actual recommended set and common recommendations. The actual recommended set is nothing but the restaurants that the user has rated as high in the original data set. This is considered as a ground truth for our recommendations. The common recommendation set is an intersection of the actual recommended set and the recommendation set obtained from the embedding. 
\par
Getting the recommended set from the embedding for a user is a step by step process. We iterate through all the top users of the data set and try and recommend restaurants from them. If a user has rated only a few restaurants as high in the original data set, then there is a high probability that the common recommendation set will be empty. To avoid this, we have put a limitation on the lower bound on the high rated restaurants set for each user. There is a limitation on the upper bound on the same set. We found the need to do this because there are some users who have rated more than hundred restaurants as high. If we take such users for recommendation, then their set will dominate in the common recommendations set. Thus, a limit helps us in reducing the bias towards a particular user along with limiting the set of the recommended restaurants from the embedding. 
\par
The next step is to predict the cluster in which the user in question lies. We pick the nearest users and their actual recommendation set is observed which form as an output of recommended restaurants from the embedding. 
While getting the restaurants of the nearest ten neighbours of the observed user from the cluster, we give weights to each restaurant as well. Let us assume we have got a list of restaurants = [$R_1, R_2, R_3$..., $R_n$]. The user in question, U, has his ten nearest neighbours has $X_1, X_2, X_3,….., X_10$. It turns out, that $R_1$ is a highly rated restaurant by both users $X_1$ and $X_5$. So, weight of $R_1$ will be 2. $R_2$ is highly rated by only $X_7$, so the weight of $R_2$ is $1$. $R_3$ belongs to the high rated restaurant sets of users $X_1, X_3, X_5, X_6 and X_8$. Therefore, its weight is 5. The resultant recommended set that is observed is in the form of a dictionary, where the restaurant is the key, and its weight is the value. For the above scenario, the output from the embedding will be like [$R_1: 2, R_2:1, R_3:5... R_n:1$]
This weight determining technique tells us which restaurant has been visited the most by the friends of the user, thus improving the quality of recommendation.

\subsection{Hybrid Recommendation}
Hybrid model can be achieved in multiple ways, using multiple available techniques. These methods have been extensively studied in \cite{kim2012hybrid} and \cite{ccano2017hybrid}. One common technique would be to concatenate embedding outputs from individual models, create a supervised data set and train a error minimizing gradient descent model. Another way would be to take outputs from each embedding and use these outputs in a hybrid filter that ranks recommendation items according to weights assigned to each model. We grossly follow the second approach in this project. Our recommendation filter takes the recommendation output from each embedding as the input to give final recommendation. Weights are learnt for each embedding using gradient descent. This collective weight then determines which restaurants are the best to recommend and thus, the resultant recommendation is calculated.
\par
We also dynamically create a data set using outputs of individual embedding and train a deep neural network to provide improved accuracy of recommendation.
Our hybrid recommendation architecture is shown in Figure 2.
\begin{figure}
\includegraphics[width=8cm,height=8cm]{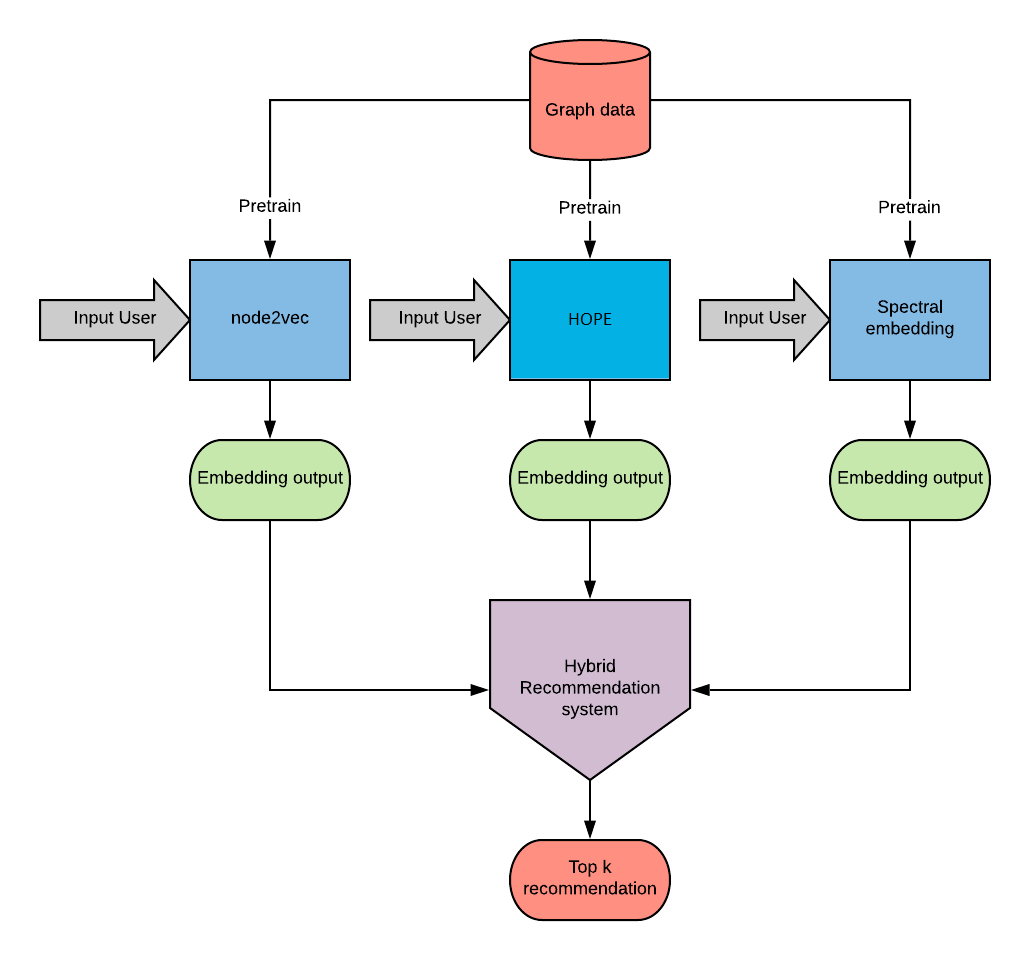}
\caption{Our Architecture for hybrid Recommendation}
\end{figure}
We build and train a deep neural network \cite{lecun2015deep} to map $X$ to $Y$ where $X$ and $Y$ are features and labels of a data set we create. This data set is created using outputs from individual embedding. Before we discuss the details of how the data set is created, we fix on something important. That is, to compare recommendation quality of hybrid model with individual embedding, we work on a constant ground by using a few selected users throughout the recommendation process. A set of 100 well connected users are chosen for whom, restaurants are recommended and MAE is calculated.
\par
We briefly discuss the hybrid data set creation process. Now that the set of users are fixed, we union all the ground truth high rated restaurants for each user which sums up to 1434. A vector of size 1434, $Y_i$ is made for each user $u_i$. In this vector $Y_i$, the ground truth restaurants for $u_i$ are represented by 1 and rest by 0. The vector $[Y_1, Y_2, .. Y_n]$ is nothing but the labels $Y$ for our data set, with the shape (100, 1434). Creating $X$ is a bit more complicated. For each user $u_i$, we create a vector $X_i$ of shape (3, 1434). Each of the 1434 entries represent one restaurant, with shape (1, 3) expressing if each embedding method (HOPE, Spectral and node2vec) could or could not recommend that restaurant to $u_i$. Hence the final shape of $X = [X_1, X_2, ... X_n]$ is (100, 3, 1434).
\par
Now, that the data set is ready, we build a deep neural network to map $X$ to $Y$. The raw architecture of the network directly taken from tensorboard is represented in Figure 3.
\begin{figure}
\includegraphics[width=8cm,height=10cm]{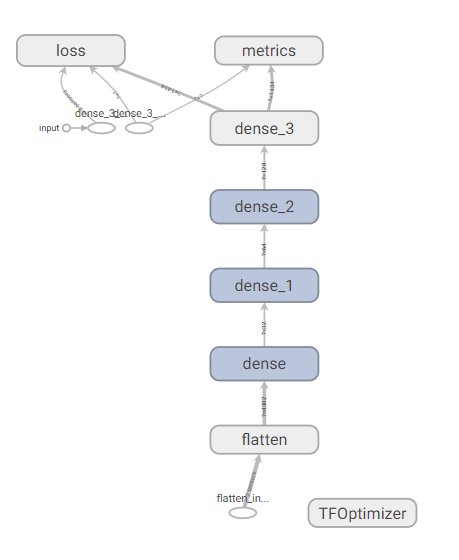}
\caption{Our Deep learning Network Architecture taken from Tensorboard}
\end{figure}
The dense layers ($dense, dense_1, dense_2$) are internal layers with (32, 64, 128) neurons respectively. $dense_3$ has the shape of $Y$, that is 1434. ReLU is used as activation function for all layers. Mean square error is taken as loss function. Adam optimizer is used with learning rate of 0.0001 and trained for 40 epochs. This number was determined by studying train and validation loss, to avoid overfitting of the network. The above configuration was found to be most optimal for the task to the best of our knowledge.

\section{\textbf{Results}}
We choose two simple evaluation metrics to determine quality and accuracy of recommendations, Mean Average Error (MAE) \cite{yang2015user} and Coverage \cite{he2010social}.
\[ MAE = \frac{1}{N} \sum \frac{|N_r - N_{hit}|}{N_r}\]
where $N$ is the size of test set, $N_r$ is number of movies recommended and $N_{hit}$ is the number of hits (correctly recommended). It evaluates the error in making recommendations. 
\[Coverage = \frac{1}{N} \sum \frac{N_{hit}}{N_u}\]
where $N_u$ is the number of movies actually rated by the user U. Coverage estimates the fraction of movies that were rated by user covered by the recommendation system. 
\par
Before going into the final metric comparisons, we compared each of the embedding technique against their MAE and Coverage values for different recommendation count. The results are as shown.

\begin{figure}
\includegraphics[width=8cm,height=6cm]{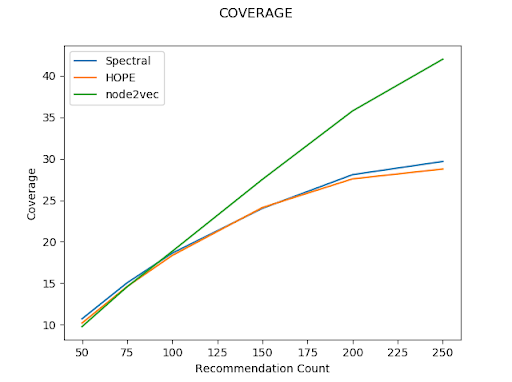}
\caption{Coverage Graph for Embedding}
\end{figure}

\begin{figure}
\includegraphics[width=8cm,height=6cm]{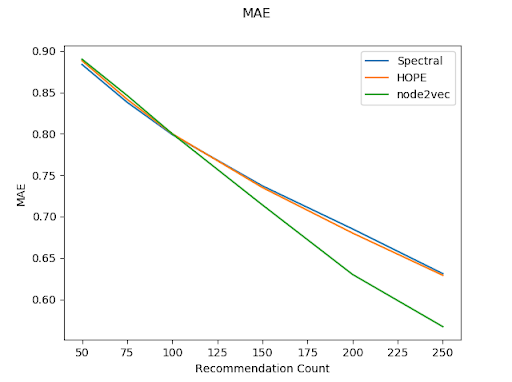}
\caption{Mean Average Error Graph for Embedding}
\end{figure}

We clearly see that node2vec performs much better than the other two equally performing embedding techniques.
\par
For a good recommendation system, MAE must be as low as possible and Coverage must be as high as possible. The results are summarized in Table ~\ref{table:comparison}.

\begin{table}
\begin{tabular}{|l|l|l|l|l|}
\hline
Recommendation Method      & \multicolumn{2}{c|}{Coverage \%} & \multicolumn{2}{c|}{MAE }                   \\ \hline
                         & Top 100       & Top 200      & Top 100                   & Top 200                \\ \hline
Spectral Clustering          & 18.62      & 28.08       &  0.799 & 0.685\\ \hline
HOPE & 18.34 & 27.57 & 0.800 & 0.689
                         \\ \hline
node2vec   & 18.84 & 35.76 & 0.801 & 0.630
                        \\ \hline
Hybrid(train)      & \multicolumn{2}{c|}{63.21} & \multicolumn{2}{c|}{0.367} \\ \hline

Hybrid(test)             &  \multicolumn{2}{c|}{48.53} &  \multicolumn{2}{c|}{0.514}
                       \\ \hline
\end{tabular}
\newline
\caption{Comparison of results from different recommendation methods}
\label{table:comparison}
\end{table}
The hybrid model entries in the table shown, are for a network trained for 40 epochs, which is gives the highest coverage for test data.

\section{\textbf{Future Work}}
There is a lot of scope for taking up the above work as a further research or into direct practical applications. We started out with an assumption that the graph is static, because recomputing embedding for a dynamic graph is much harder. Efficient dynamic graph embedding generation for social graphs can be taken up as a research problem. A few work has been done on this by \cite{nguyen2018continuous} , \cite{goyal2018dyngem} and \cite{li2017attributed}.
\par
More graph embedding as  discussed in \cite{cai2018comprehensive} and \cite{goyal2018graph} can be explored and performance can be compared on other social graph data sets. In the above results, there is a huge gap between train and test results, resulting from the inherent problem in social recommendation task i.e. even though friends influence our decision, individual user at times makes his own random decision and chooses items. Such issues can be taken into consideration in future work. Also the novel hybrid approach proposed relies on a dynamically created training data directly created from the outputs of individual pipeline. Though this method is effective at finding a pattern among users and their interests, training data is created at every run with new set of users. More robust and advanced hybrid approaches discussed in \cite{ccano2017hybrid} can be used for item recommendation. 
\par
Implicit item to item graph can be created based on item similarity. Embedding can be applied to this graphs, which can improve recommendations. Similarly, embedding can be generated from heterogeneous user to item graphs as done in \cite{liu2019real}. These graphs are more informative and would definitely improve the task of recommendation.

\section{\textbf{Conclusion}}
We started out by proposing that social circles affect the choices of a user in a social environment. Hence, users would mostly buy an item, watch a movie or go to a restaurant if their close friends recommended the items, movies or restaurants. This property of social behaviour is mathematically represented by graph embedding that cluster similar nodes nearby in a given n-dimensional space.
\par
We then used multiple graph embedding techniques on social graphs to compare performances on item recommendation task. We filtered out three embedding techniques i.e. Node2Vec, HOPE and spectral clustering for our social graph. We found that Node2Vec performed the best among all three for recommendation task.
\par
We also started with a hypothesis that a hybrid of multiple embedding would yield better results as each embedding focuses on  multiple aspects of the graph. This hypothesis is validated in the end.  We found that hybrid model does perform considerably better than other individual embedding proving our hypothesis to be correct.

\bibliographystyle{ieeetr}
\bibliography{main}

\end{document}